\documentclass[sigconf]{acmart}

\usepackage{booktabs} 
\usepackage{eurosym}
\usepackage{xcolor}
\usepackage{multirow}
\setcopyright{acmcopyright}

\usepackage{standalone}
\usepackage{tikz}
\usetikzlibrary{positioning}
\usepackage{soulutf8}

\renewcommand\footnotetextcopyrightpermission[1]{} 
\begin{document}
\title{Data Economy for Prosumers in a Smart Grid Ecosystem}

\author{Ricardo J. Bessa}
\orcid{0000-0002-3808-0427}
\affiliation{%
  \institution{INESC TEC}
  \streetaddress{Campus da FEUP, Rua Dr. Roberto Frias}
  \city{Porto, Portugal} 
  \postcode{4200-465}
}
\email{ricardo.j.bessa@inesctec.pt}

\author{David Rua}
\affiliation{%
  \institution{INESC TEC}
  \streetaddress{Campus da FEUP, Rua Dr. Roberto Frias}
  \city{Porto, Portugal} 
  \postcode{4200-465}
}
\email{drua@inesctec.pt}

\author{Cl\'audia Abreu}
\affiliation{%
  \institution{INESC TEC and FEUP}
  \streetaddress{Campus da FEUP, Rua Dr. Roberto Frias}
  \city{Porto, Portugal} 
  \postcode{4200-465}
  }
\email{claudia.r.abreu@inesctec.pt}

\author{Paulo Machado}
\affiliation{%
  \institution{INESC TEC}
  \streetaddress{Campus da FEUP, Rua Dr. Roberto Frias}
  \city{Porto, Portugal} 
  \postcode{4200-465}
}
\email{paulo.a.machado@inesctec.pt}

\author{Jos\'e R. Andrade}
\affiliation{%
  \institution{INESC TEC}
  \streetaddress{Campus da FEUP, Rua Dr. Roberto Frias}
  \city{Porto, Portugal} 
  \postcode{4200-465}
}
\email{jose.r.andrade@inesctec.pt}

\author{Rui Pinto}
\affiliation{%
  \institution{INESC TEC and FEUP}
  \streetaddress{Campus da FEUP, Rua Dr. Roberto Frias}
  \city{Porto, Portugal} 
  \postcode{4200-465}
}
\email{rui.b.pinto@inesctec.pt}

\author{Carla Gon\c{c}alves}
\affiliation{%
  \institution{INESC TEC and FCUP}
  \streetaddress{Campus da FEUP, Rua Dr. Roberto Frias}
  \city{Porto, Portugal} 
  \postcode{4200-465}
}
\email{carla.s.goncalves@inesctec.pt}

\author{Marisa Reis} 
\affiliation{%
 \institution{INESC TEC and FEUP}
 \streetaddress{Campus da FEUP, Rua Dr. Roberto Frias}
 \city{Porto, Portugal}
 \postcode{4200-46}}
 \email{marisa.m.reis@inesctec.pt}

\renewcommand{\shortauthors}{R.J. Bessa et al.}

\begin{abstract}
Smart grids technologies are enablers of new business models for domestic consumers with local flexibility (generation, loads, storage) and where access to data is a key requirement in the value stream. However, legislation on personal data privacy and protection imposes the need to develop local models for flexibility modeling and forecasting and exchange models instead of personal data. This paper describes the functional architecture of an home energy management system (HEMS) and its optimization functions. A set of data-driven models, embedded in the HEMS, are discussed for improving renewable energy forecasting skill and modeling multi-period flexibility of distributed energy resources.    
\end{abstract}


\maketitle

\section{Introduction}

The deployment of smart meters at the domestic prosumer level is enabling emergent regulated and non-regulated data-based services that can help to boost distributed energy resources (DER) integration and promote customer empowerment \cite{McGranaghan2016}. 

At the regulated level, the ongoing discussion about the future roles of Distribution System Operators (DSO) is focused in smart metering data management (data manager role) and electricity market facilitation (neutral market facilitator role) \cite{Mallet2014, CEER2016}. In this context, the following paragraphs discuss examples of recent initiatives within the smart grid ecosystem.

The European Union (EU) project UPGRID developed and demonstrated a Neutral Market Access Platform (NMAP) and a Retailer Platform (RP). The NMAP is hosted by a DSO and encompasses the exchange of information, including consumption profiles from the DSO and flexibility profiles from the Home Energy Management System (HEMS) \cite{Alonso2017}. The RP is responsible for receiving information or requests from the UPGRID platform, process and send this information to its HEMS. The EU project FLEXICIENCY developed and demonstrated a pan-European Market Place that aims at delivering services and exchange of data, tools, methodologies, in a standardized way across Europe \cite{Stromsather2015}. The platform receives/submits data/service request and readdresses requests to the DSO and service providers platforms where the data, services, software and tools are located \cite{Boukir2017}. Green Button (U.S.A.) is an industry-led work that provides a common format for electrical energy metering data so that electricity consumers can access their data in an easily readable and secure format via a ``Green Button'' on their electric utilities' website. Once customers access their data, they can share it as they choose, by independent choice and action, with those they trust. Third parties (e.g., energy retailer, energy services company) services are also emerging with this initiative, e.g. a solar developer could use customer-metering data to determine optimal system size with a more accurate cost-benefit analysis \cite{Gerza2015}. A review of data management models in eight European countries can be found in \cite{CEER2016}. 

In terms of non-regulated services or third-party services, several works in the literature describe data-driven services and business cases to boost demand response potential and promote energy efficiency. Smart meter data can be used by a third-party for segmentation of customers and identification of temporal consumption patterns \cite{Chelmis2015}, predicting customer response to price signals \cite{Kwac2016}, estimating the price elasticity of customers \cite{Gomez2012} and derive optimal bidding strategies under dynamic electricity tariffs \cite{Song2017}. In fact, total energy consumption at the residential level is enough to derive a ranking of thermal-load flexibility and sub-metering is not required \cite{Albert2018}.

Smart meter and HEMS data can also be used to induce behavioral changes in energy consumption \cite{Tsuda2017}, for instance through gamification techniques \cite{Gustafsson2009} or broadcast of information (e.g., price signal) \cite{Eksin2018}. In \cite{McKenna2012}, a detailed analysis of the requirements for different applications of smart meter data (e.g., balancing, demand response, network planning) is conducted considering the European legislation about personal data protection.  

The General Data Protection Regulation (GDPR) approved by the EU Parliament on 14 April 2016, harmonizes data privacy laws across Europe, protect and empower all EU citizens data privacy \cite{GDPR}. The new regulation introduces a new fine (i.e., up to 4\% of annual global turnover or 20 M\euro), intelligible and easily accessible form for data access consent, right to access data and be forgotten (erase personal data and cease further dissemination of the data). Moreover, it requires ``privacy by design'', which means inclusion of data protection from the start in systems designing, rather than an addition. Therefore, in order to build data economy or data marketplaces focused on electrical energy consumption data, it is necessary to design knowledge extraction methods that ensure data privacy from the beginning.     

Presently, some researchers are starting to design demand response algorithms that avoid sharing and transfer of personal data. For instance, data-driven pricing schemes for load shifting without access to personal load requirements \cite{Xu2018} or information exchange models where consumers keep their load levels private and participate in a real-time price scheme \cite{Eksin2018}. 

In this context, the present paper presents the Horizon 2020 InteGrid project's vision for the HEMS and its embedded intelligent functions. The core goal is to design forecasting algorithms (load, solar, flexibility) keeping data local and private and at the same time create economic value for stakeholders such as retailers and aggregators. The paper starts by describing the functional architecture, hardware and software integration of the HEMS in section \ref{sec:HEMS}. Then, proposes distributed forecasting services in section \ref{sec:forecast} and flexibility forecast and modeling of behind the meter energy resources in section \ref{sec:flexibility}. The potential for future work is discussed in section \ref{sec:conclusions}.

\section{Home Energy Management System}\label{sec:HEMS}

The HEMS concept was initially introduced as a central unit located within a domestic building with the capability of performing an optimized control of behind the meter energy resources  \cite{livengood_phd_2011}. This optimization was loosely defined as a cost reduction procedure, typically achieved through a lower cost use of energy considering internal factors (e.g., devices and system that are able to provide energy use flexibility) and external factors (e.g. weather conditions, discriminated energy prices). HEMS can be exploited as a platform that is able to support demand side management schemes allowing a generalized participation of consumers in such services \cite{Williams_2012}.

The main characteristic of current HEMS implementations is the ability to allow the monitoring of energy consumption by means of existing metering devices and a user interface that allows the representation of data in a user-oriented way (e.g. dashboards). Another characteristic is the ability to automate existing resources and by using very simple control mechanisms allow the operation of devices and systems at more convenient hours (e.g. pre-program load activation to hours in which the energy costs are lower). The use of sensor technology for remote control of appliances based on a threshold or an user-defined activations allows exploiting energy savings potential. Optimization schemes are currently being sought to allow the optimization of existing flexible energy resources according to multiple criteria and multiple restrictions.

To enhance the capability of fully optimizing the energy use in households, HEMS have been incorporating energy models of devices, systems and spaces so that their specific characteristics along with user preferences can in fact be properly considered when defining their optimal activation \cite{Ozturk_2013}. Users are thus able to insert configurations and preferences and take advantage of local flexibility to activate their energy resource at an optimal time. This involves a planning stage and a time-ahead operation, being an example the case where the end-user sets the preferences and configurations, an optimal operation schedule is determined, and in the next day the existing resources are activated according to the schedule.

\subsection{Functional Architecture}

\begin{figure}[hbt]
	\centering
	\includegraphics[scale=.58]{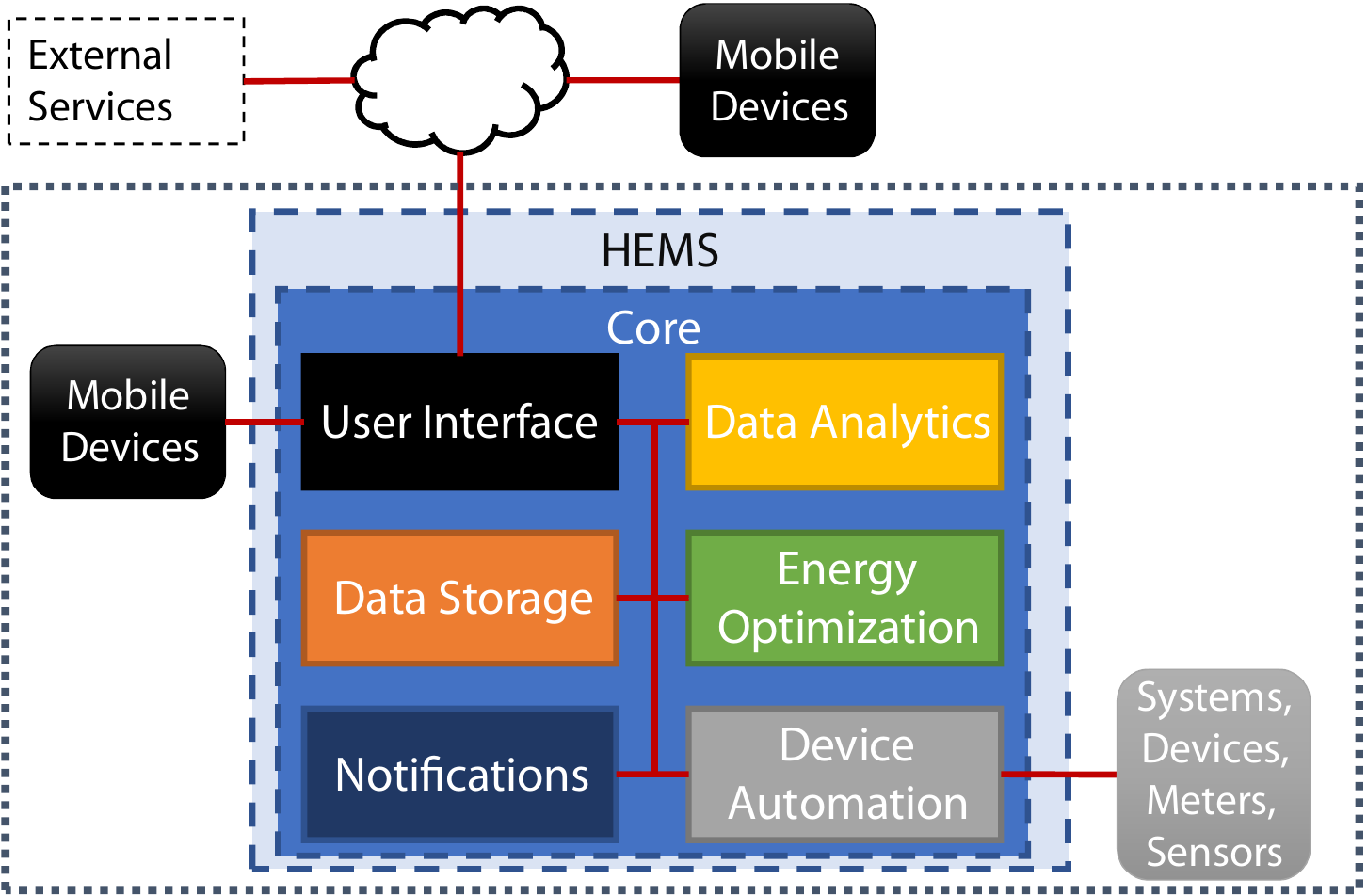}
	\caption{HEMS functional architecture.}
	\label{fig:hems_arch}
\end{figure}

HEMS are currently flexible and modular HW and SW platforms, capable of supporting a wide variety of features and functionalities the allow end-users to take advantage of existing incentives and define a customizable operation schedule. In Fig.~\ref{fig:hems_arch} there is a representation of the architecture of the HEMS under design for InteGrid project.

The functionalities of a HEMS are typically grouped into: user interface, energy optimization and automation. User interface (UI) is responsible for presenting data and relevant information to the user and allow the insertion of configurations and preferences. The energy optimization creates an energy representation of the household, through the parameterization of the energy models of devices and systems according to their configuration, and according to the selected criteria establish a time-ahead operation schedule. The automation takes care of the communication between the HEMS central system and the associated devices and systems to retrieve monitoring and operation state data and to set remote configurations or actions that allow an optimal schedule to be implemented.

Other groups can be defined to provide additional features to HEMS: data analytics, data storage and event notifications. Data analytics provides data handling functions related to forecasting and machine learning to extract context information about the availability of energy resources and likelihood of specific energy consumption patterns. Section~\ref{sec:forecast} describes in detail the forecasting features. Data storage functionalities ensure secure data storage and availability to mobile terminals that interact with the HEMS both locally and remotely. Event notification manages alarms and announces unexpected events according to different priority schemes, allowing end-users to received filtered information. They also allow end-users to use non-smart loads, optimally, by receiving that information with a configurable time-ahead.

\subsection{Optimization Strategies}

One of the distinguishable feature that an HEMS must provide is the capability of computing optimal (or suboptimal) energy use schedules that provide added-value to the end-users. There are different optimization strategies that can be exploited through a HEMS and they largely depend on the criterion (or set of criteria) that might be established. Typically, a cost reduction benefit is sough, and that can be achieved either through energy consumption reduction or through an optimal load allocation considering energy consumption prices. While the former is associated to behavioral changes the latter is more related to the technical aspects of the energy management. In \cite{Abreu2017}, an HEMS implementation, based on the work carried out in the AnyPLACE project, is presented, with the mathematical formulation of optimization strategies for cost reduction considering the variable and fixed components of the energy consumption costs.

\begin{figure}[hbt]
	\centering
	\includegraphics[scale=.26]{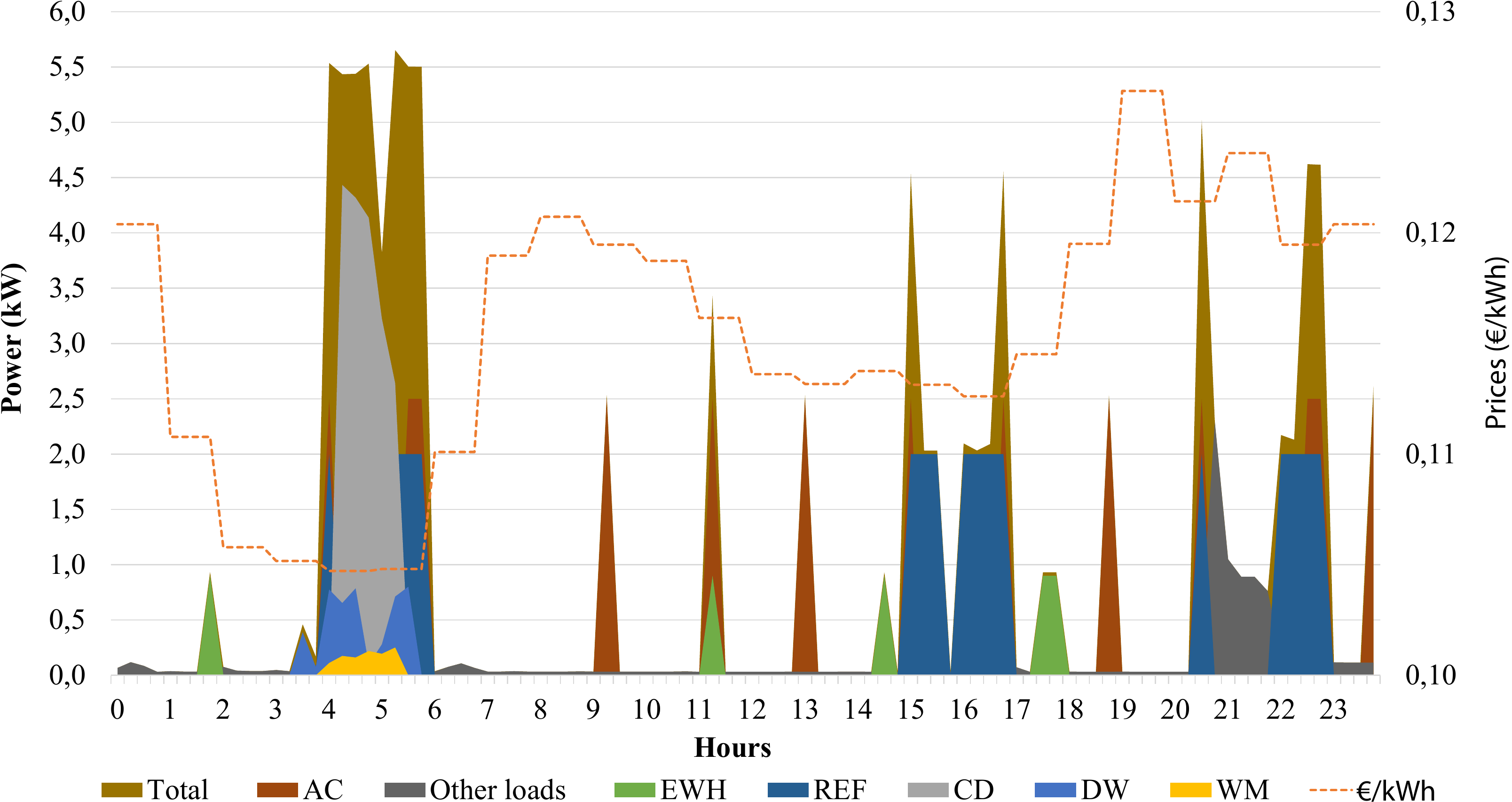}
	\caption{Example of price-based optimization.}
	\label{fig:price_opt}
\end{figure}

The HEMS being designed for InteGrid project is able to optimize the energy use considering variable price tariffs or the local production of PV. As presented in Fig. \ref{fig:price_opt} loads can be allocated to time periods in which the price is lower to allow a reduction on energy expenditures.

\begin{figure}[hbt]
	\centering
	\includegraphics[scale=.28]{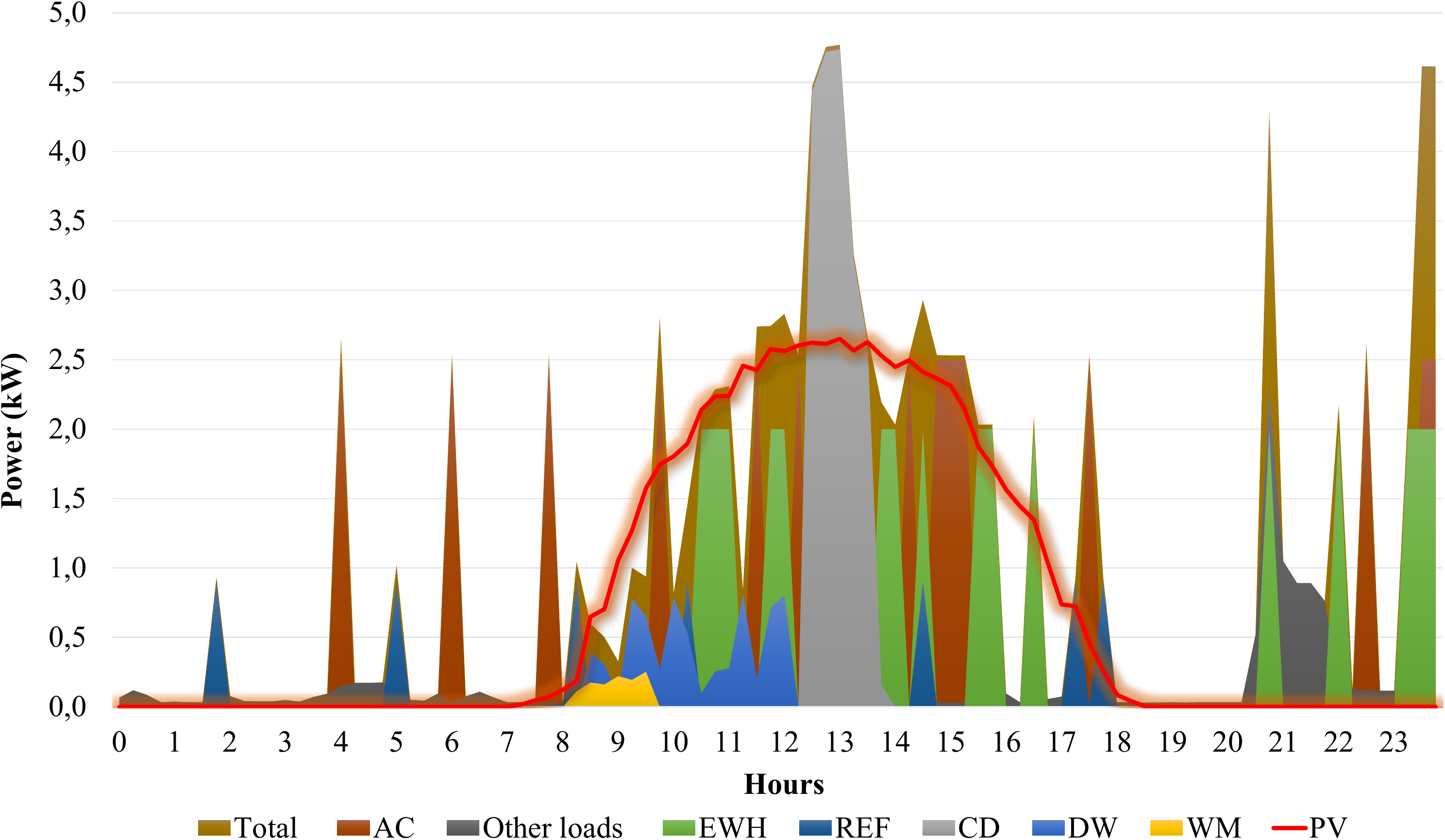}
	\caption{Example of PV self-consumption optimization.}
	\label{fig:pv_opt}
\end{figure}

When considering PV microgeneration, controllable appliances are shifted toward PV production (according to the forecast generated in the previous day), as portrayed in Fig. \ref{fig:pv_opt}.

Thermal loads such as air conditioners (AC) and Electric Water Heaters (EWH) have specific energy models whereas dishwashers (DW), washing machines (WM) and cloths dryers (CD) are modeled as shiftable loads. These are modeled as average power and average operating time either input by the end-user or based on estimated load consumption profiles.

One of the objectives of InteGrid is to make use of a fully representable energy model for devices, systems and spaces to allow a HEMS a higher observability and controllability with positive impact in the quality of the optimization process. 

\subsection{Hardware Integration}

One key aspect of a HEMS is the ability to integrate existing hardware as well as new devices and systems upon which it can leverage to produce more and more complex energy optimization schemes that exploit conveniently their specific technical characteristic and the users' preferences. 

The support of these existing or new systems provide a significant improvement on the user experience and comfort by the reduction of manual inputs needed that can be directly obtained through these systems. Also, since the accuracy of those inputs tend to be higher, the quality of the results is improved.

This underlying modular and interoperable nature addresses also the fact that not all the devices and systems that the HEMS may manage are smart or automated, considering means of integrating older devices through user engagement, by means of friendly notifications (email or other) inviting the users to actively control the devices at home, or through the use of other devices like smart plugs that can be integrated to provide some sort of automation.

In order to keep implementation costs low and at the same time ensure a robust computational solution a Raspberry Pi 3 single board computer is used. The use of Linux allows the use of OpenHAB as an open automation support platform that makes use of binding implementations to exchange information with different devices and systems. Despite there is a considerable number of manufacturers providing bindings for building automated devices it is possible to develop binding implementations to support specific devices, which means that virtually any device can be integrated within the HEMS platform.

\section{Forecasting Services}\label{sec:forecast}

\subsection{Concept}

The works in \cite{Tastu2014, Bessa2015, Cavalcante2017} showed that geographically distributed time series data can improve the renewable energy forecasting skill up to six lead-times ahead. Time series data from different sources or owners are combined in a vector autoregression (VAR) model and the LASSO penalty structure is used to uncover sparse structures in the coefficients matrix. The VAR process captures linear interdependencies among multiple time series, enabling each HEMS to model its PV time series evolution based not only on its own lagged values but also in the lagged values collected by other HEMS.  The main challenge is on how to combine data from multiple HEMS owners and maintain data privacy, which is analyzed from the mathematical point of view in subsection \ref{sec:col_for}.  

Furthermore, information from a spatial grid of Numerical Weather Predictions (NWP) is valuable to improve renewable energy forecast accuracy for multiple days \cite{Andrade2017}. Here, the challenge is to process the NWP grid data and run the statistical forecasting algorithms locally at the HEMS level. Subsection \ref{sec:for_rap} presents computational results of forecasting algorithms (section \ref{sec:HEMS}) running in the HEMS hardware.

\begin{figure}[hbt]
	\centering
	\includegraphics[scale=.3]{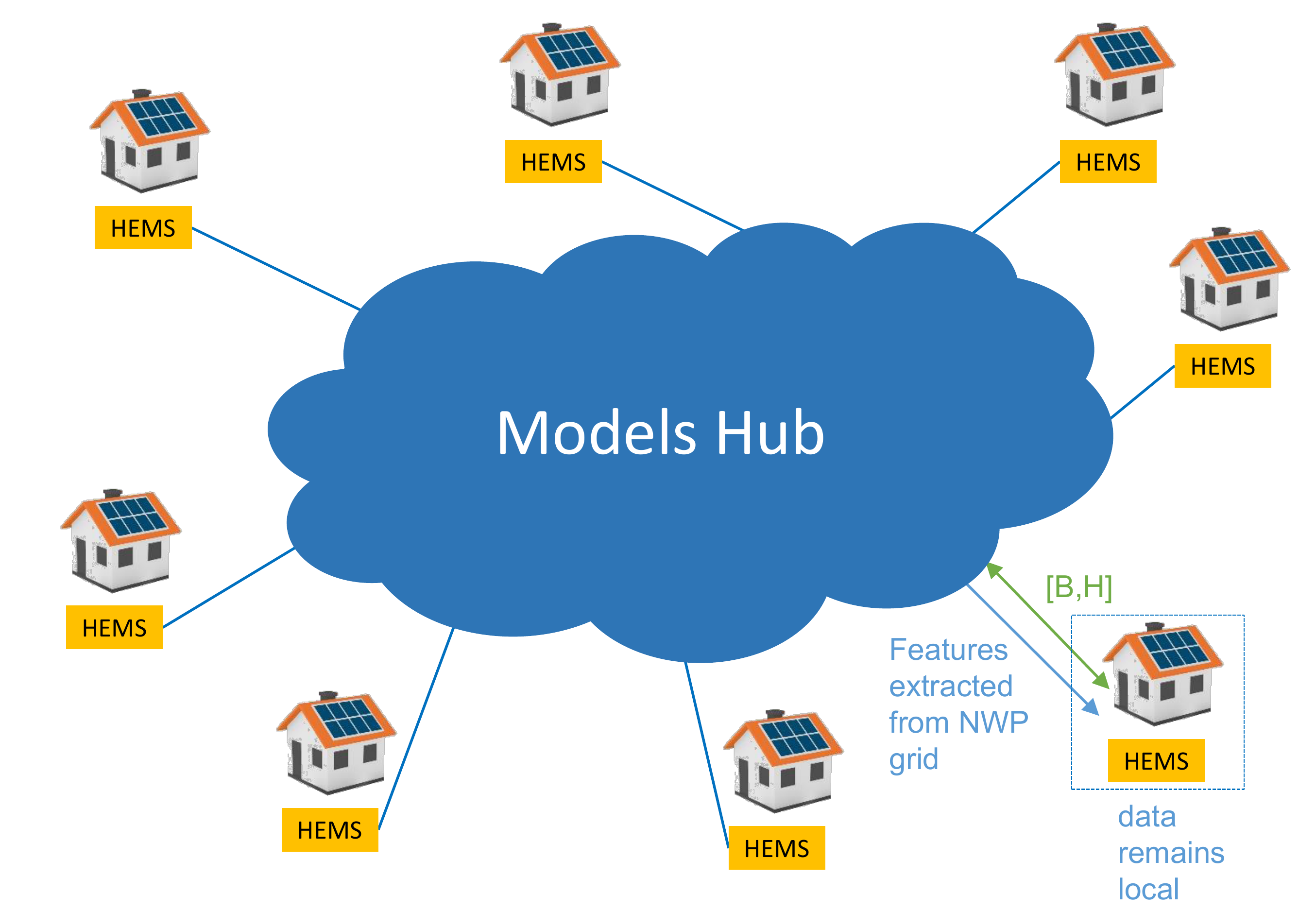}
	\caption{Models hub concept for renewable energy forecasting.}
	\label{fig:models_hub}
\end{figure}

Figure \ref{fig:models_hub} illustrates the models hub concept that serves two goals: (a) central node in a collaborative forecasting scheme (i.e. VAR model) where each HEMS exchanges the matrices from the B and H-update steps of the Alternating Direction Method of Multipliers (ADMM) method \cite{Boyd2011}; (b) apply feature engineering techniques that extract information from the NWP grid and send the post-processed variables to the HEMS where a embedded gradient boosting tree algorithm is used to produce solar power forecasts. 

\subsection{Embedded Forecasting}\label{sec:for_rap}

Gradient boosting trees (GBT) algorithm combined with feature engineering techniques capable of extracting information from a spatial grid of NWP and resuming it in a new smaller subset of variables can led to substantial improvements on point and probabilistic forecasts for short-term horizons (i.e., up to 72 hours ahead) \cite{Andrade2017}. However, the high quantity of information to process represents a major constraint and makes it prohibitive to deploy in small computational units such as HEMS. Table~\ref{tb:rpi3_desktop_specs} presents a comparison between the technical specifications of the HEMS and a conventional desktop computer.

\begin{table}[hbt]
	\centering
	\renewcommand{\arraystretch}{1.15}
	\caption{Computational resources comparison between the HEMS and a conventional desktop computer.}
	\label{tb:rpi3_desktop_specs}
	\begin{tabular}{|c|c|c|}
		\hline
		Comp. & Raspberry Pi 3 & Desktop \\ \hline
		\multirow{3}{*}{CPU}	& Broadcom BCM2837	& Intel(R) Core(TM)\\
        							& ARM Cortex-A53, 1.2GHz	& i7-6700, 3.4GHz \\
                                    & 4 cores, 4 threads & 4 cores, 8 threads \\
                                    \hline
		\multirow{1}{*}{RAM}	& 1GB LPDDR2@900MHz	& 12 GB DDR4@2133MHz\\
                                \hline
		OS & Linux-Raspbian & Windows 10 \\ \hline
		
	\end{tabular}
\end{table}

Table~\ref{tb:rpi3_desktop_specs} reveals that the RAM size and processing power of the HEMS are a clear limitation when trying to process large quantities of NWP data in an acceptable computational time. For this reason, the envisioned solution uses a centralized distributed computing framework, nested in the Models Hub platform, to process the raw NWP grid data and create relevant features from the NWP grid. These variables, combined with NWP forecasts for the client location, feed a GBT model that runs locally in each HEMS.

For the HEMS, the challenge of embedded forecasting is summarized in four phases:
\begin{itemize}
	\item Initial request to retrieve historical post-processed NWP grid variables from the Models Hub platform, for the timespan of historical observed PV data.
	\item Apply feature engineering techniques to extract temporal information from NWP variables for the HEMS location.
	\item Fit the GBT model using the historical of observed PV generation values and respective NWP variables. The model is updated on a daily or weekly basis.
	\item Generate PV forecasts based on the NWP variables for an horizon up to 48 hours ahead and the post-processed operational forecasts requested to the Models Hub platform.
\end{itemize}

\subsubsection{Accuracy Results Analysis}

In this subsection, improvements on the PV forecasting skill resulting from the introduction of spatial-temporal features extracted from the NWP grid are demonstrated. The metrics Mean Absolute Error (MAE) and Root Mean Squared Error (RMSE) are considered to assess the point forecast quality. The Continuous Ranked Probability Score (CRPS) is used to evaluate the quality of probabilistic forecasts. An extended description of the feature engineering process can be found in \cite{Andrade2017}.

The importance of variables extracted from the NWP grid is here evaluated by comparing two models that contain extra temporal and spatial information with a base reference model exclusively composed by seasonal and NWP variables for the grid central point. Table~\ref{tb:meteo_variables_tbl} describes the input information of the base model.

\begin{table}[hbt]
	\renewcommand{\arraystretch}{1.1}
	\caption{Base model input variables.}
	\label{tb:meteo_variables_tbl}
	\begin{tabular}{|c|l|}
		\hline
		Type & Variables\\
		
		\hline
		\multirow{2}{*}{Seasonal}	& Month of the year\\ 
									& Hour of the day\\ 
		\hline
		\multirow{5}{*}{NWP} 		& swflx [W/m$^2$] - shortwave flux\\ 	
									& temp [K] - ambient temperature (2 meters)\\
									& cfl [0,1] - cloud cover at low levels\\
									& cfm [0,1] - cloud cover at mid levels\\
									& cfh [0,1] - cloud cover at high levels\\
									& cft [0,1] - cloud cover at low and mid levels\\               
		\hline
	\end{tabular}
\end{table}

The following two models accommodate the new information. Model T is representative of the information that can be extracted from the NWP series for the HEMS location, and Model F comprises a selection of the best spatial and temporal variables that were able to maximize the forecast skill.
\begin{itemize}
	\item \textbf{Model T} - Temporal information extracted from the NWP series to the HEMS location:
	\begin{itemize}
		\item Lags and leads.
		\item Temporal variance with centered windows of 3h, 7h, 11h.
		\item Information from different NWP runs.
	\end{itemize}
	\item \textbf{Model F} -  Spatial information extracted from the NWP grid: 
	\begin{itemize}
		\item Hourly spatial standard deviation of NWP grid variables.
		\item Hourly spatial weighted average of NWP grid variables.
		\item Principal components applied individually to the grid information of each NWP variable mentioned in Table~\ref{tb:meteo_variables_tbl}.
	\end{itemize}
\end{itemize}

The performance of these models is evaluated over a timespan of two years (from May 1$^{st}$ 2015 to 28$^{th}$ June 2016) with a sliding window of 12 months. Considering $\epsilon$ as a metric score, and \textit{base} as a the base model, the improvement of a model is given by $\left(1 - \frac{\epsilon_{m}}{\epsilon_{base}}\right) \times 100$ (\%).

\begin{figure}[hbt]
	\centering
	\includegraphics[scale=0.38]{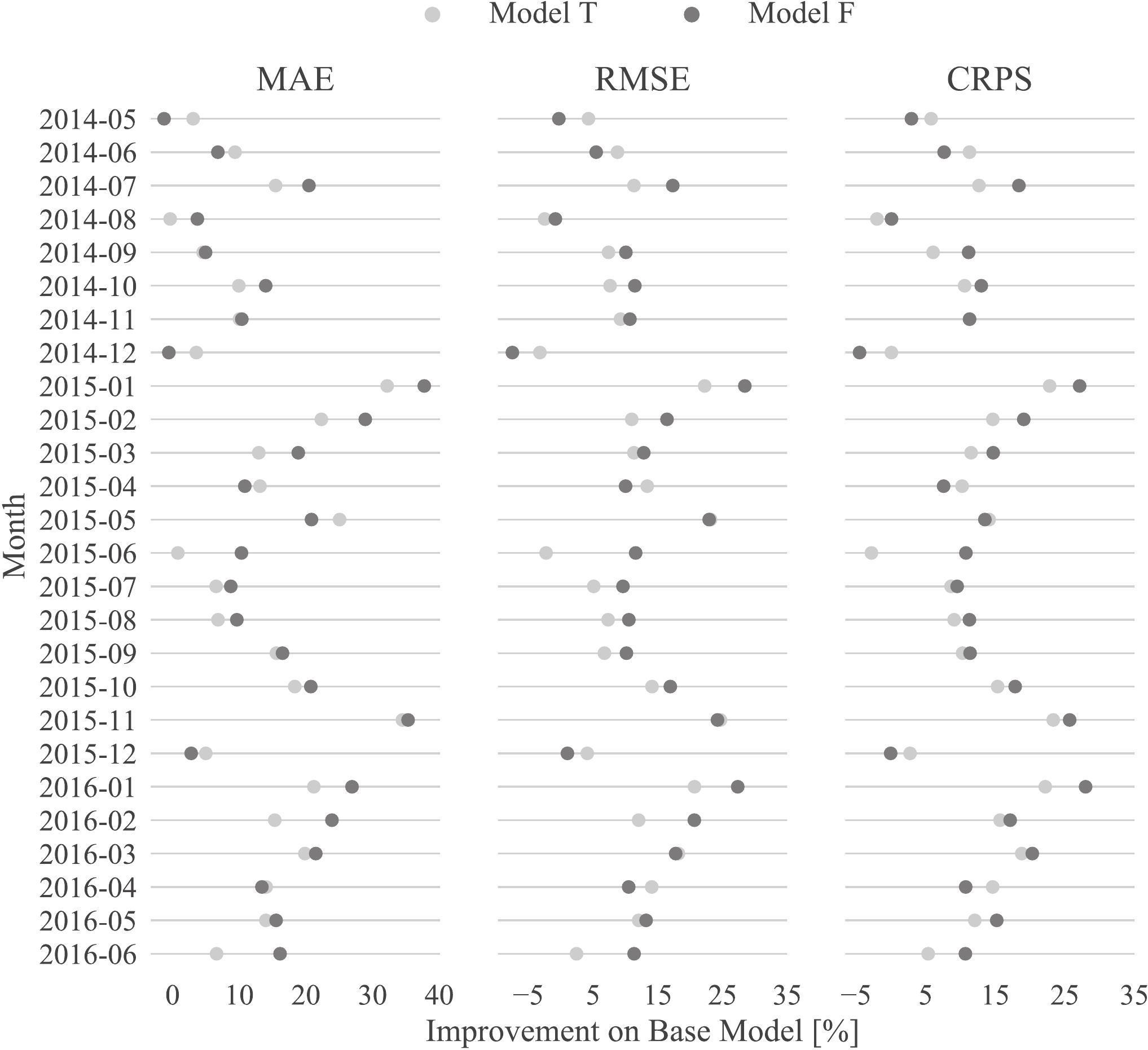}
	\caption{Monthly relative improvements of models T and F over the base model.}
	\label{fig:pv_forec_models_imp}
\end{figure}

Figure~\ref{fig:pv_forec_models_imp} depicts the average improvements of every aforementioned model over the reference base model for forecasts within an horizon of 24 hours ahead. The evaluation is computed on out-of-sample datasets, guaranteeing that in each fold the training data is not contaminated by test samples. The night periods were removed for the PV generation.

An analysis of the figure shows that the two new models greatly outperform the base model in every month of the evaluation period. By itself, the collection of temporal variables (Model T) already provides great average improvements of 13.73\%, 10.35\%, 11.33\% on MAE, RMSE and CRPS that peak in November 2015 at 34.48\%, 24.67\%, 23.32\%. However, the maximum forecast skill is only obtained by including the spatial information (Model F), which led to average improvements of 16.09\%, 12.85\%, 13.11\% for the same metrics, that peak at 37.75\%, 28.43\%, 27.97\% in January 2015.

As final remark, it is important to underline that a good forecast quality can be achieved by solely depending on the temporal information extracted from the NWP runs for the HEMS location. However, to maximize point and specially probabilistic forecasts quality, the NWP grid information is necessary.

\subsubsection{Computational Times Evaluation}

Although the forecasting algorithms successfully run in the Raspberry Pi 3, a significant impact on the computational time of each model run is verified when compared to a conventional desktop computer. It is important to underline that the GBT regression model requires a separated training for each quantile of the probabilistic forecast. Table \ref{tb:computational_times} shows the total computational times necessary to train all the regression models and to compute forecasts for the two systems described in Table~\ref{tb:rpi3_desktop_specs}.

\begin{table}[hbt]
	\centering
	\renewcommand{\arraystretch}{1.1}
	\caption{Comparison of computational time results.}
	\label{tb:computational_times}
	\begin{tabular}{|c|c|c|}
		\hline
		Device			& GBT Fitting & Operational Forecast		\\
		\hline
		Desktop			& 42s			& 0.03s					\\ 
		Raspberry Pi 3	& 320s			& 0.49s		\\
		\hline
	\end{tabular}
\end{table}

\subsection{Collaborative Forecasting} \label{sec:col_for}

\subsubsection{VAR-LASSO model and ADMM}

Here, a brief review of the VAR-LASSO model is presented, as well as the ADMM formulation for distributed parameter estimation. 

Let $y_{i,t}$ be the time series of an $i$--th HEMS, in time $t$, and $\{\mathbf Y_t\}=\{(y_{1,t},\dots,y_{n,t})\}$ a $n$-dimensional vector time series. Then, a VAR model of order $p$ describes the trajectory of $\mathbf Y_t$ as 
\begin{equation} \label{eq:VARp}
	\mathbf Y_t=\boldsymbol \eta + \sum_{\ell=1}^p \mathbf B^{(\ell)} \mathbf Y_{t-l} + \boldsymbol \varepsilon_t,
\end{equation}
where  $(\mathbf B^{(\ell)})_{i,j}$ represents the parameters for time series $i$, associated with lag $\ell$ of time series $j$; 
 $\boldsymbol\eta=(\mu_{1},\dots,\mu_{n})^T$ is the vector of constant terms, 
 and $\boldsymbol\varepsilon_t=(\varepsilon_{1,t},\dots,\varepsilon_{n,t})^T$ is a white noise term. By simplification, $\mathbf Y_t$ is assumed to be a centered process, i.e., $\boldsymbol\eta=\mathbf 0$.
 
In order to formulate a matrix representation of VAR(p) model, let $\mathbf Y=(\mathbf Y_1,\dots,\mathbf Y_T) \in \mathbb{R}^{(n\times T)}$ the response matrix, $\mathbf B=(\mathbf B^{(1)},\allowbreak\dots,\allowbreak\mathbf B^{(p)})  \in \mathbb{R}^{(n\times np)}$, the coefficient matrix, $\mathbf Z=(\mathbf Z_1,\dots,\mathbf Z_T)  \in \mathbb{R}^{(np\times T)}$ the explanatory variables matrix, with $\mathbf Z_t=(\mathbf Y_{t-1},\dots,\allowbreak\mathbf Y_{t-p})  \in \mathbb{R}^{(n\times T)}$, and $\mathbf E=(\boldsymbol\varepsilon_1,\dots,\boldsymbol\varepsilon_T)\in \mathbb{R}^{(n\times T)}$ the error matrix.
Then, 
\begin{equation} \label{eq:matricialVARp}
	\mathbf Y=\mathbf{BZ+ E}.
\end{equation}

Commonly, the system (\ref{eq:matricialVARp}) is solved computing the coefficients that minimize $\|\mathbf{Y-BZ}\|_2^2$, where $\|.\|_r$ will represent both vector and matrix $L_r$ norms.  However, high-dimensional data can introduce irrelevant or redundant information making convenient the application of the LASSO framework, which is a regularized version of least squares that introduces an $L_1$ penalty on the coefficients. In standard VAR-LASSO approach, the coefficients are estimated by
\begin{equation} \label{eq:lasso}
	\arg \min_{\mathbf B} \Big(\frac{1}{2}\|\mathbf Y-\mathbf{BZ}\|_2^2+\lambda \|\mathbf B\|_1\Big),
\end{equation}
where $\lambda > 0$ is a scalar penalty parameter.

Since the cost function in~(\ref{eq:lasso}) is non-differentiable, the ADMM constitutes a powerful algorithm to solve this problem, making it possible to perform a parallel optimization. Summarily, the ADMM rewrites the VAR-LASSO objective function~(\ref{eq:lasso}) replicating the $\mathbf B$ variable using the $\mathbf H$ variable ($\frac{1}{2}\|\mathbf{Y -BZ}\|_2^2+\lambda\|\mathbf H\|_1$) and adding an equality constraint imposing $\mathbf{B=H}$. 
Then, based in the augmented Lagrangian of this reformulated minimization problem (with penalty parameter $\rho>0$), the ADMM formulation of (\ref{eq:lasso}) consists in the following iterations~\cite{Cavalcante2017}:
\begin{equation} \label{ADMMLASSOVAR} \left
		\{
            \begin{array}{l}
              \mathbf B^{k+1} :=\text{argmin}_{\mathbf B}\Big( \frac{1}{2}\|\mathbf{Y -BZ}\|_2^2+\frac{\rho}{2}\|\mathbf{B}- \mathbf H^k+\mathbf U^k\|_2^2\Big) \\
              \mathbf H^{k+1}:=\text{argmin}_{\mathbf H}\Big(\lambda \|\mathbf H\|_1 + \frac{\rho}{2}\|\mathbf B^{k+1}-\mathbf H+\mathbf U^k\|_2^2\Big)\\
              \mathbf U^{k+1}:=\mathbf U^k+\mathbf B^{k+1}-\mathbf H^{k+1}.
            \end{array}
          \right.
\end{equation}

Given that $\|\mathbf{Y-BZ}\|_2^2$ and $\|\mathbf H\|_1$ are decomposable, the minimization problem over $\mathbf B$ and $\mathbf H$ can be separately solved for distributed data. Therefore, the ADMM provides a desirable formulation for parallel computing. 

\begin{figure}[bt]
\def\bsize{0.52}
\def\ncol{1}
\def\adjs{-0.25}

\begin{tikzpicture}
\small

\draw (-4,-1.3) rectangle (-4+3*\bsize,-1.3+\ncol) node[pos=0.5] (B) {$\text{B}$};
\draw (-2,-1.3+\adjs) rectangle (-2+\ncol,-1.3+3*\bsize+\adjs) node[pos=0.5] (Z) {$\text{Z}$};
\draw  (-4.2,1.1) rectangle (-0.6,-2) node[pos=0.5,yshift=1cm] (BN1) {$\text{Main formulation}$};
 
\draw[fill=blue!60] (0,0) rectangle (\bsize,\ncol) node[pos=0.5] (B11) {$\text{B}_1$};
\draw (\bsize,0) rectangle (2*\bsize,\ncol) node[pos=0.5] (p) {$\dots$};
\draw[fill=blue!20] (2*\bsize,0) rectangle (3*\bsize,\ncol) node[pos=0.5] (BN1) {$\text{B}_N$};

\draw[fill=blue!60] (4*\bsize,2*\bsize+\adjs) rectangle (4*\bsize+\ncol,3*\bsize+\adjs) node[pos=0.5] (Z11) {$Z_1$};
\draw (4*\bsize,1*\bsize+\adjs) rectangle (4*\bsize+\ncol,2*\bsize+\adjs) node[pos=0.5] (p) {$\dots$};
\draw[fill=blue!20] (4*\bsize,0*\bsize+\adjs) rectangle (4*\bsize+\ncol,1*\bsize+\adjs) node[pos=0.5] (ZN1) {$Z_N$};
\draw  (-0.4,1.6) rectangle (3.5,-0.35) node[pos=0.5,yshift=1cm] (BN1) {};
\draw (-2.3,1.1) |- (1.4,1.9) node[pos=0.5,xshift=1.8cm,yshift=0.25cm] (B) {$\text{Splitting across predictors}$ -- equation system~(\ref{eq:admmrow})};
\draw[->, >=latex] (1.4,1.9) -- (1.4,1.6);

\draw[fill=blue!60] (0,-1) rectangle (3*\bsize,-1-\ncol/3) node[pos=0.5] (B11) {$\text{B}_1$};
\draw (0,-1-1*\ncol/3) rectangle (3*\bsize,-1-2*\ncol/3) node[pos=0.5] (p) {$\dots$};
\draw[fill=blue!20] (0,-1-2*\ncol/3) rectangle (3*\bsize,-1-3*\ncol/3) node[pos=0.5] (B11) {$\text{B}_N$};

\draw[fill=blue!60] (4*\bsize,-1-0*\ncol/3-\adjs) rectangle (4*\bsize+\ncol/3,-1-3*\bsize-\adjs) node[pos=0.5] (Z11) {$Z_1$};
\draw (4*\bsize+\ncol/3,-1-0*\ncol/3-\adjs) rectangle (4*\bsize+2*\ncol/3,-1-3*\bsize-\adjs) node[pos=0.5] (Z11) {$\dots$};
\draw[fill=blue!20] (4*\bsize+2*\ncol/3,-1-0*\ncol/3-\adjs) rectangle (4*\bsize+\ncol,-1-3*\bsize-\adjs) node[pos=0.5] (Z11) {$Z_N$};
\draw (-2.3,-2) |- (1.5,-2.8) node[pos=0.5,xshift=1.8cm,yshift=-0.25cm] (B) {$\text{Splitting across examples}$ -- equation system~(\ref{eq:admmcolumn})};
\draw[->, >=latex] (1.5,-2.8) -- (1.5,-2.5);
\draw  (-0.4,-2.5) rectangle (3.5,-0.6) node[pos=0.5,yshift=1cm] (BN1) {};

\end{tikzpicture}
\caption{Two main scenarios for distributed computation.}
\label{scenariossummary}
\end{figure}
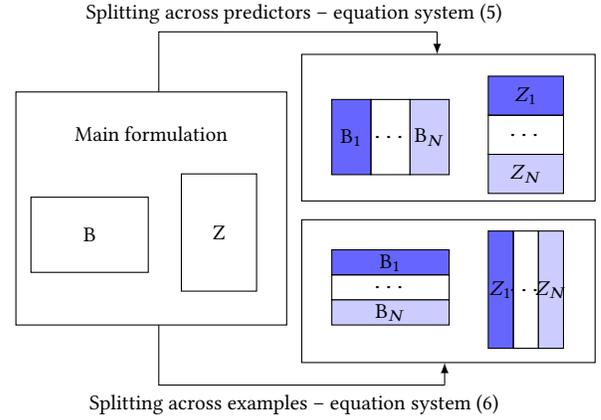

Figure~\ref{scenariossummary} illustrates the two most commonly used approaches to split the optimization problem, in which $\mathbf Z$ is partitioned into $N$ row blocks (splitting across predictors) or $N$ column blocks (splitting across examples). Conventionally, this two generic formulations are called Consensus Optimization and Sharing Optimization, respectively. The corresponding ADMM formulation using~(\ref{ADMMLASSOVAR}) is presented in the systems of equations~(\ref{eq:admmrow}) and (\ref{eq:admmcolumn}),
\begin{align} \small \label{eq:admmrow}\left
		\{
            \begin{array}{l}
              \mathbf B_i^{k+1} :=\text{argmin}_{\mathbf B_i}\Big( \frac{\rho}{2}\|\mathbf B_i^k \mathbf Z_i+\overline{\mathbf H}^k-\overline{\mathbf{BZ}}^k-\mathbf U^k-\mathbf B_i \mathbf Z_i\|_2^2+ \\
              \multicolumn{1}{r}{\lambda\|\mathbf B_i\|\Big)}\\
              \overline{\mathbf H}^{k+1}:=\frac{1}{N+\rho}\Big(\mathbf Y+\rho \overline{\mathbf{BZ}}^{k+1}+\rho \mathbf U^k\Big)\\
              \mathbf U^{k+1}:=\mathbf U^k+\overline{\mathbf{BZ}}^{k+1}-\overline{\mathbf H}^{k+1}
            \end{array}
          \right.
\end{align}
\begin{align} \small \label{eq:admmcolumn} \left
		\{
            \begin{array}{l}
              \mathbf B_i^{k+1} :=\text{argmin}_{\mathbf B_i}\Big( \frac{1}{2}\|\mathbf Y_i+ \mathbf B_i\mathbf Z_i\|_ 2^2+\frac{\rho}{2}\|\mathbf B_i-\mathbf H^k+\mathbf U_i^k\|_2^2\Big) \\
              \mathbf H^{k+1}:=S_1\Big(\overline{\mathbf B}^{k+1}+\overline{\mathbf U}^k,\frac{\lambda}{N\rho}\Big)\\
              \mathbf U_i^{k+1}:=\mathbf U_i^k+\mathbf B_i^{k+1}-\mathbf H^{k+1}
            \end{array}
          \right.
\end{align}
\normalsize
where $\overline{\mathbf{BZ}}^k=\frac{1}{N}\sum_{j=1}^N \mathbf B_j^k \mathbf Z_j$, $\overline{\mathbf{B}}^k=\frac{1}{N}\sum_{j=1}^N \mathbf B_j^k$ and $S_1$ is the scalar soft thresholding operator, defined as $S_1(x,a)=\frac{x}{|x|}\max\{0,|x|-a\}$. 

\subsubsection{Can ADMM ensure data privacy in collaborative forecasting?}

The main challenge is to investigate the potential of these iterative systems of equations to develop a collaborative forecasting with data privacy preserved. However, as will be concluded, the direct application of these two alternatives cannot fulfill the privacy requirement. 

If the problem is divided across predictors, then $\hat{\mathbf Y}=\sum_i \mathbf B_i \mathbf Z_i$ and each HEMS will estimate $\mathbf B_i$ using $\mathbf Z_i$, which is composed uniquely by the lags of its own time series. 
Hence, each HEMS computes the corresponding $\mathbf B_i$ without share data. The problem is the update of $\mathbf H$ where each HEMS should share its $\mathbf B_i \mathbf Z_i + \mathbf (\mathbf 0, \mathbf Y_i,\mathbf 0)$, allowing the more curious HEMS to recover $\mathbf Y$.

On the other hand, if the problem is divided across examples, then each $\mathbf B_i$ (computed in parallel) is estimated using $\mathbf Z_i$, which in this case consists of specific lags of all time series, meaning data is shared between all participants. Nevertheless, this division by examples can be interesting for single-output problems in the energy sector, where private data from multiple consumers is needed.

Notice that in both situations, the existence of a neutral element may be imposed, through which all HEMS communicate, i.e., assuming a centralized process so that HEMS do not exchange results directly. However, this would imply that the neutral element can reconstruct the original data. Furthermore, if in equation~(\ref{eq:admmrow}) the neutral element provides the $i$--th HEMS $\overline{\mathbf H}^k$,$\overline{\mathbf{BZ}}^k$ and $\mathbf U^k$, then $i$--th HEMS can reconstruct $\mathbf Y$.

Ideally, the algorithm would be adapted in a way that nobody can reconstruct the original data, be they HEMS or the neutral element. Even better, the algorithm should be decentralized (i.e. peer-to-peer), asynchronous (to cope with delays or communication failures) and time-adaptive (enabling the assimilation of new samples as they become available in order to improve the model with time, updating the coefficients of the model, without starting from scratch). 

In the literature, related works may be found. A decentralized structure with asynchrony and delays was proposed in~\cite{wu2016decentralized}, in which the workers can communicate independently with their neighbors, at different times and for different durations. In terms of online ADMM, a general approach may be explored in~\cite{hosseini2014online}. Latter work proposed an asynchronous time-adaptive ADMM for consensus optimization~\cite{matamoros2017asynchronous}.

With respect to algorithms that take into account data privacy: the authors of~\cite{zhang2017privacy} proposed a privacy-preserving decentralized (but neither asynchronous nor online) optimization method, introducing time-varying penalty matrices on ADMM method and using partially homomorphic cryptography. 

\section{Surrogate Models for Energy Flexibility} \label{sec:flexibility}

This section describes three surrogate representations of behind the meter energy resources flexibility. The main idea is to exchange flexibility models estimated in the HEMS, instead of behind-the-meter data, which can be take the form of: set of flexibility trajectories, machine learning model or dynamic virtual batteries. It is important to stress that other representations are possible, such as simplified models to quantity thermal-based demand response potential and the HEMS only sends upstream the estimated model and not the consumption data \cite{Albert2018}.

\subsection{Set of Flexibility Trajectories} 

HEMS can include diverse equipment such as electric water heaters (EWH), domestic small-scale batteries, photovoltaic (PV) panels, and HVAC systems. The flexibility that HEMS can offer relates to deviations from the net-load profiles that were expected to occur, and has its explanation based on the flexible nature of the previously enumerated HEMS equipment. 

The concept of flexibility trajectory embraces the multi-period HEMS flexibility potential instead of independent single-periods formulation \cite{Pinto2017} . A feasible flexibility trajectory represents the HEMS potential of reshaping its expected net-load profile while guaranteeing technical operation viability of all equipment and internal constraints such as EWH water temperature, battery state-of-charge, or even room air-temperature. Additionally, user-defined constraint regarding the use of the battery can be also implemented (e.g., aiming at minimizing the energy spilling during PV generation surplus). The uncertainty in the net-load profile was also modeled, namely by considering different PV generation scenarios. Ultimately, a feasible trajectory must comply with the defined constraints for a pre-establish percentage of PV generation scenarios (e.g. feasible in 90\% of the PV scenarios).

\begin{figure}[hbt]
	\centering
	\includegraphics[scale=.32]{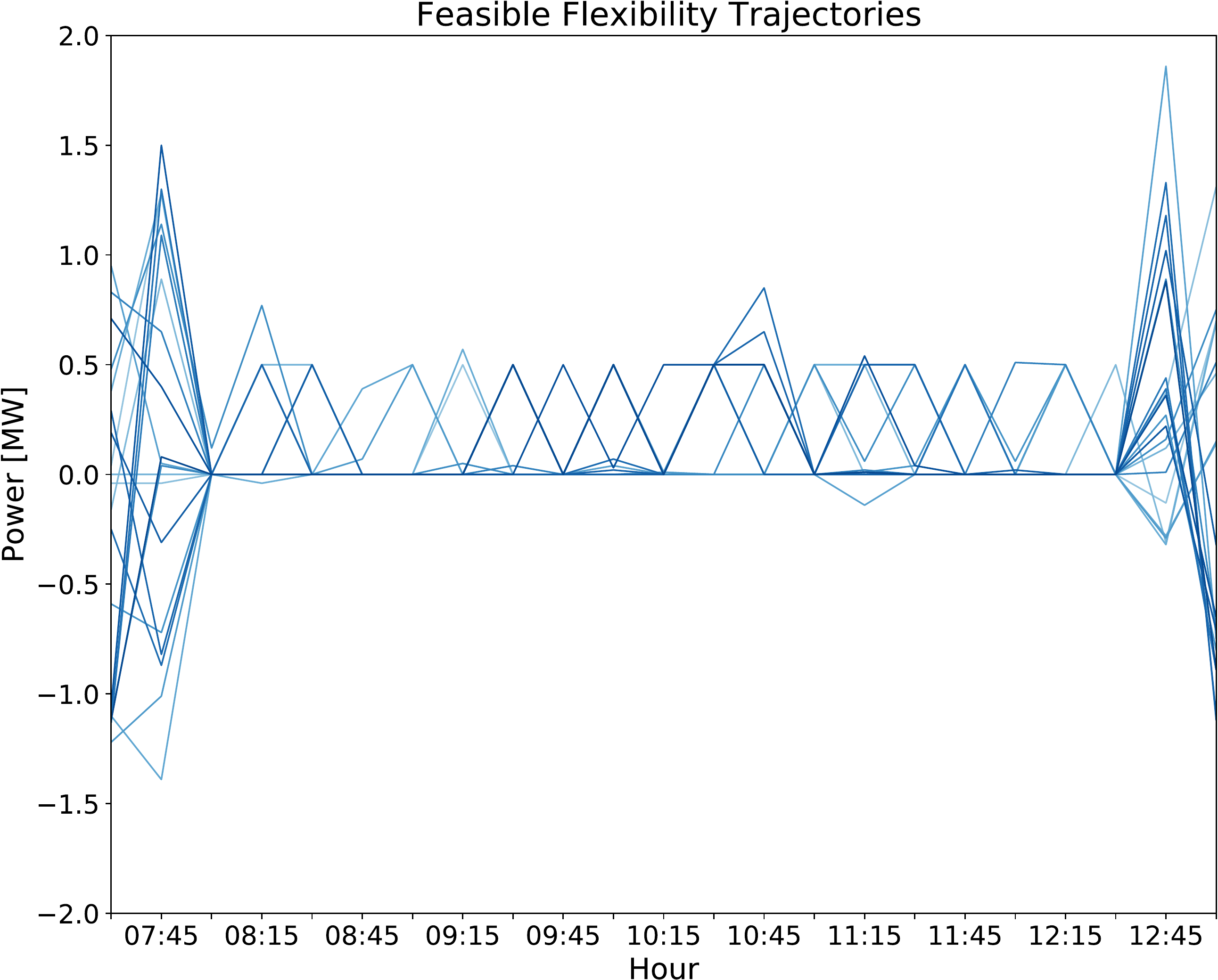}
	\caption{Illustrative set of 20 EPSO-generated feasible flexibility trajectories.}
	\label{fig:trajectories}
\end{figure}

Defining the feasible domain for the HEMS flexibility potential is a complex task. An Evolutionary Particle Swarm Optimization (EPS) based algorithm was proposed in \cite{Pinto2017} to search and sample the mentioned feasible domain. The final output is a set of feasible trajectories that can be used to describe the HEMS multi-temporal flexibility potential. Figure \ref{fig:trajectories} depicts a set of 20 feasible flexibility trajectories encompassing a period where PV generation surplus occurs.

\subsection{Machine Learning Model}

The set of flexibility trajectories described in the previous section can be learned (or presented) by a support vector data description (SVDD) algorithm \cite{Bremer2011,Pinto2017}. SVDD is a machine learning one-class support vector machine that can be used to classify ``feasible'' or ``unfeasible'' HEMS operating trajectories upon request from an optimization or control algorithm.

In SVDD, the flexibility trajectories set ($X$) is summarized with a combination of support vector ($x_i$) and respective coefficients ($\beta_i$). This describes a high-dimension sphere delimiting the feasible domain. A new trajectory ($x$) is classified by comparing the sphere radius with its radius calculated as follows:  

\begin{equation} \label{eq:svdd}
	R^{2}(x)=1-2\sum_{i}\beta\textsubscript{i}k(x\textsubscript{i}, x) + \sum_{i,j}\beta\textsubscript{i}\beta\textsubscript{j}k(x\textsubscript{i}, x\textsubscript{j})
\end{equation}

where $R^2$ is the square of the radius being calculated, $x_i$ and $x_j$ are support vectors, $k$ is the kernel function. For this problem, it was found that sigmoid kernel is the most suitable type \cite{Pinto2017}.
To be classified as feasible, a trajectory must present a radius lower or equal to the sphere's radius.
The results for the test case described in \cite{Pinto2017} are presented in Table \ref{tb:performance}.

The main limitation of this representation is that it only allows to classify trajectories as feasible or unfeasible and interpretability (in terms of energy flexibility) is low. However, Eq. \ref{eq:svdd} can be easily integrated in optimization or control problems and maintain personal data private. 

\subsection{Dynamic Virtual Battery}

An alternative to the SVDD representation for the set of flexibility trajectories is a virtual battery model (similar to the one used in \cite{Hao2015}). The virtual battery representation might be seen as a linear system resembling the operation of a battery parameterized by power limits for charge (\textit{P\textsubscript{t}\textsuperscript{max}}) and discharge (\textit{P\textsubscript{t}\textsuperscript{min}}) cycles, by maximum (\textit{SOC\textsubscript{t}\textsuperscript{max}}) and minimum (\textit{SOC\textsubscript{t}\textsuperscript{min}}) limits for the state of charge (SOC), and also by the initial SOC level (\textit{SOC\textsuperscript{ini}}).

The power and SOC limits modeled in this work are dynamic throughout the time horizon that is being considered, instead of using a fixed value for all the periods considered. This way, the feasible domain represented by the virtual battery will be more adapted regarding the period-based information coming from the set of feasible flexibility trajectories.

The problem formulation for the virtual battery model aims at minimizing the battery size (SOC), as well as minimizing, for each period, the power amplitude represented by the difference between the charge  and discharge limits. Equations \ref{eq:f_obj}-\ref{eq:power_limiti} summarize the problem formulation, where \textit{traj\textsubscript{k, t\textsubscript{i}}} refers to the power value represented by the trajectory \textit{k} from the flexibility set for the time period \textit{t\textsubscript{i}}, and \textit{T} refers to the number of periods considered.

\begin{equation} \label{eq:f_obj}
	min \enspace \sum_{t=1}^{T} SOC\textsubscript{t}\textsuperscript{max} + \sum_{t=1}^{T} SOC\textsubscript{t}\textsuperscript{min} + \sum_{t=1}^{T} P\textsubscript{t}\textsuperscript{max} - \sum_{t=1}^{T} P\textsubscript{t}\textsuperscript{min}
\end{equation}

\begin{equation} \label{eq:soc_limiti}
	SOC\textsubscript{t\textsubscript{i}}\textsuperscript{min} <= SOC\textsuperscript{ini} + \sum_{t=1}^{t\textsubscript{i}} traj\textsubscript{k, t\textsubscript{t}} <= SOC\textsubscript{t\textsubscript{i}}\textsuperscript{max},\enspace \forall t\textsubscript{i}\in[1, T] \enspace, \enspace \forall k
\end{equation}

\begin{equation} \label{eq:power_limiti}
	P\textsubscript{t\textsubscript{i}}\textsuperscript{min} <= traj\textsubscript{k, t\textsubscript{i}} <= P\textsubscript{t\textsubscript{i}}\textsuperscript{max},\enspace \forall t\textsubscript{i}\in[1, T] \enspace and \enspace \forall k
\end{equation}

Compared to the SVDD representation, the virtual battery model can be easily integrated in a optimal power flow formulation, interpreted by an end-user in terms of available flexibility and it is scalable for prosumers aggregation algorithms.

Regarding the performance on correctly classifying trajectories, the virtual battery model shows a better performance  compared to the SVDD model when classifying feasible trajectories. On the other hand, for unfeasible trajectories the virtual battery presents a higher number of wrongly classified trajectories. Table \ref{tb:performance} presents the percentage of correctly classified trajectories for both models on two different sets: feasible and unfeasible trajectories. Results refer to an 8-hours time horizon.

\begin{table}[]
\centering
\caption{Performance: SVDD vs. Virtual Battery.}
\label{tb:performance}
\begin{tabular}{c|c|c|}
\cline{2-3}
                                      & Feasible traj. & Unfeasible traj. \\ \hline
\multicolumn{1}{|c|}{SVDD}            & 95 \%                 & 80 \%                     \\ \hline
\multicolumn{1}{|c|}{Virtual Battery} & 100 \%                & 70 \%                     \\ \hline
\end{tabular}
\end{table}\label{tab:traj}

\section{Conclusions}\label {sec:conclusions}

The increasing awareness of prosumers about data protection and business value of their smart meter data demands for a revision of the current data management paradigms and business use cases. Platforms like HEMS allow consumers to implement energy efficiency actions through automation based on optimal scheduling of existing energy resources. The HEMS are a source of a considerable amount of data and it can include several modules and support different functionalities (e.g., data analytics) to deal with complex data structures.

Forecasting algorithms of renewable energy and net-load benefit from exchange of information between peers (e.g., geographically distributed observations), which shows significant challenges in maintaining data private. A recent trend for mobile devices is the Federated Learning concept\footnote{\url{https://research.googleblog.com/2017/04/federated-learning-collaborative.html} (access on February 2018)} that keeps all training data on the device and only the statistical or machine learning model is shared.  Only the updated model is sent to the cloud, using encrypted communication. Furthermore, technologies like TensorFlow Lite for mobile and embedded device developers will enable machine learning tasks in data that does not leave the device and with faster local computation. The advantages are clear, no dependency on network connection and training with less data. However, collaborative forecasting with data privacy remains an open area of research.

The aggregation of flexibility from prosumers also demands for a model-based approach with local computations to extract flexibility parameters. In this context, the interpretability of the virtual battery model is appealing . However, several challenges arise for model-driven representation: inclusion of forecast uncertainty in the flexibility quantification; multi-temporal nature of the flexibility activation; combination of different flexible resources. This type of models have relevant information to implement economic demand response schemes and it is possible to create a marketplace for such models. For instance, virtual batteries flexibility can be traded through the combination of blockchain and smart contracts \cite{Mattila2016} or integrated in a distributed optimal power flow (OPF) problem. 

An interesting initiative is the OpenMined\footnote{\url{https://openmined.org/} (accessed on February 2018)} deep learning marketplace that combines federated machine learning, blockchain, multi-party computation, and homomorphic encryption. In this ecosystem, deep learning algorithms fitted in distributed data blocks are traded through smart contracts. The SVDD parameters of a flexibility set or the virtual batteries parameters can be traded in a similar platform.

The future research directions are: (a) a non-linear peer-to-peer vector autoregression model with data fully private and that can be applied to different collaborative forecasting problems; (b) aggregation of virtual battery models that combine prosumer's flexibility and forecast uncertainty and its integration with the HEMS hardware. 

In the framework of the InteGrid project, the HEMS and its intelligent functions will be installed in real prosumers in Portugal and Sweden and the value of the proposed approaches will be assessed in a real-world scenario.

\section*{Acknowledgements}

The research leading to this work is being carried out as a part of the InteGrid project (\textit{Demonstration of INTElligent grid technologies for renewables INTEgration and INTEractive consumer participation enabling INTEroperable market solutions and INTErconnected stakeholders}), which received funding from the European Union's Horizon 2020 Framework Programme for Research and Innovation under grant agreement No. 731218, and AnyPLACE (\textit{Adaptable Platform for Active Services Exchange}), which received funding from the European Union's Horizon 2020 Framework Programme for Research and Innovation under grant agreement No. 646580.

\bibliographystyle{ACM-Reference-Format}
\bibliography{ref_ACM} 

\end{document}